\documentclass[a4paper,10pt,twoside]{cpc-hepnp}

\usepackage{multicol}
\usepackage{graphicx}
\usepackage{booktabs}
\usepackage{amssymb,bm,mathrsfs,bbm,amscd}
\usepackage[tbtags]{amsmath}
\usepackage{lastpage}
\usepackage{CJK}
\usepackage{lineno}

\begin{document}

\fancyhead[c]{\small Submitted to ¡®Chinese Physics C¡¯ 2015 - 0101} \fancyfoot[C]{\small 2015 - 0101 \, -\,\thepage \,-}

\footnotetext[0]{Received XX XX 201X}

\title{Measurement of the fluorescence quantum yield of bis-MSB\thanks{Supported by National Natural Science Foundation of China (11205183,11225525,11390381) }}

\author{
DING Xue-Feng$^{1,2;1)}$\email{XF.Ding@whu.edu.cn}
\quad WEN Liang-Jian$^{2}$
\quad ZHOU Xiang$^{1;2)}$\email{xiangzhou@whu.edu.cn (corresponding author)}
\quad DING Ya-Yun$^{2}$
\quad YE Xing-Chen$^{1;2}$
\\
\quad ZHOU Li$^{2}$
\quad LIU Meng-Chao$^{2}$
\quad CAI Hao$^{1;3)}$\email{hcai@whu.edu.cn (co-corresponding author)}
\quad CAO Jun$^{2}$
}
\maketitle

\address{%
$^1$ Hubei Nuclear Solid Physics Key Laboratory, Key Laboratory of Artificial Micro- and Nano-structures of Ministry of Education, and School of Physics and Technology, Wuhan University, Wuhan 430072, China\\
$^2$ Institute of High Energy Physics, Chinese Academy of Sciences, Beijing 100049, China\\
}

\begin{abstract}
The fluorescence quantum yield of bis-MSB, a widely used liquid scintillator wavelength shifter, was measured to study the photon absorption and re-emission processes in liquid scintillator. The re-emission process affects the photoelectron yield and distribution, especially in a large liquid scintillator detector, thus must be understood to optimize the liquid scintillator for good energy resolution and to precisely simulate the detector with Monte Carlo. In this study, solutions of different bis-MSB concentration were prepared for absorption and fluorescence emission measurements to cover a broad range of wavelengths. Harmane was used as a standard reference to obtain the absolution fluorescence quantum yield. For the first time we measured the fluorescence quantum yield of bis-MSB up to 430 nm as inputs required by Monte Carlo simulation, which is 0.926$\pm$0.053 at $\lambda_{\rm ex}$ = 350 nm.
\end{abstract}



\footnotetext[0]{\hspace*{-3mm}\raisebox{0.3ex}{$\scriptstyle\copyright$}2014
Chinese Physical Society and the Institute of High Energy Physics
of the Chinese Academy of Sciences and the Institute
of Modern Physics of the Chinese Academy of Sciences and IOP Publishing Ltd}%

\begin{multicols}{2}

\section{Introduction}
Liquid scintillator (LS) is widely used in neutrino detectors. A ternary LS was adopted by the Daya Bay Reactor Neutrino Experiment~\cite{Dayabay_PRL,Dayabay_Det,Dayabay_LS}, doped with gadolinium in the target region and undoped in the gamma catcher region that surrounds the target region.  The LS consists of the solvent Linear Alkyl Benzene (LAB), the primary solute 2,5-diphenyloxazole (PPO) (3 g/L), and the secondary solute p-bis-(o-methylstyryl)-benzene (bis-MSB) (15 mg/L). The same LS recipe (undoped) will be used for the Jiangmen Underground Neutrino Observatory (JUNO)~\cite{JUNO}.

When a charged particle deposits energy in the LS via ionization, LAB molecules are excited first since the solute is only a tiny fraction of the LS. The excitation energy transfers from the LAB molecules to the PPO molecules via non-radiative processes involving the dipole-dipole interaction and short-distance collisions. The excited PPO molecules then emit photons at high scintillation efficiency when de-exciting. Many liquid scintillators only contain solvent and primary solute. To reduce the self-absorption of the scintillation light in the LS, or to shift the wavelength of the scintillation light to the optimal sensitive region of the photomultiplier tube (PMT), a secondary solute is often added. The secondary solute absorbs the photons emitted by the primary solute, then re-emits at a longer wavelength. It is therefore also called a wavelength shifter. In general, LS absorption at long wavelengths is much smaller than at short wavelengths. As a consequence, more photons can arrive at the PMTs and increase the photo-statistics, which is critical for a large neutrino detector such as JUNO to achieve the required high energy resolution.

The light re-emission is isotropic in the LS and has a certain quantum efficiency. Therefore, it affects the photoelectron yield and distribution. It also affects the photoelectron contribution of the Cerenkov light produced together with the scintillation by the charged particle. Most Cerenkov light is emitted at short wavelengths and will be absorbed and re-emitted in the LS. In a low energy neutrino experiment, many charged particles, e.g. Compton scattering electrons produced by $\gamma$s, have low energy close to (or lower than) the threshold of Cerenkov light emission. Therefore, the Cerenkov light contribution varies with the particle energy and introduces energy non-linearity to the detector. The re-emission efficiency will further modify the fraction of  Cerenkov light and impacts on the energy non-linearity~\cite{Dayabay_shape,ZhangFH_NL,YangMS_NL}. The re-emission process must be understood to optimize the LS recipe for good energy resolution and to produce a precise Monte Carlo simulation of the detector.

The fluorescence quantum yield is defined as the ratio of the number of emitted photons to the number of absorbed photons. Re-emission could happen in the molecules of all LS components, i.e. the solvent LAB, the primary solute PPO, and the secondary solute bis-MSB in the Daya Bay and JUNO LS. The solvent LAB has very low fluorescence emission efficiency, and energy is mostly transferred to the PPO via non-radiative processes. The absorption band of the wavelength shifter is always chosen to overlap with the PPO emission spectrum. Therefore, the absorption and re-emission happening in the bis-MSB is dominant. Since  re-emission is an optical process, the fluorescence quantum yield of bis-MSB can be studied by exciting the bis-MSB molecule directly with ultraviolet-visible (UV-vis) light. In general the quantum yield should be independent of the wavelength of the excitation light (aka excitation wavelength) when it is shorter than the emission wavelengths of bis-MSB, because the fluorescence photons come dominantly from the de-excitation of the first single excitation state to the ground state. When the excitation wavelength is in the middle of the emission spectrum, the solvent-solute interaction will play an important role~\cite{RedEdgeReview}, and the quantum yield could depend on the excitation wavelength.

In the literature, the fluorescence quantum yield of bis-MSB was measured to be 0.94 in cyclohexane~\cite{Argon} and 0.96$\pm$0.03 in LAB~\cite{XiaoHL} at relatively short excitation wavelength. However, the quantum yield at longer excitation wavelength, when it is in the middle of the emission spectrum and the absorption re-emission probability in the liquid scintillator starts to decrease, is of more interest to understand photon propagation in the detector. Such measurements are difficult because the absorption of the excitation light is too small and the self-absorption of the re-emitted light is large. In this study we measure the fluorescence quantum yield of bis-MSB from 345 nm up to 430 nm for the first time, with multiple samples of different bis-MSB concentrations.

\section{Measurement method}

The fluorescence quantum yield is defined as
\begin{equation}
\label{relativeFLQY_p}
\varphi(\lambda_{\rm ex}) = \dfrac{N_{\rm emitted}}{N_{\rm abs}(\lambda_{\rm ex})}\,,
\end{equation}
where $N_{\rm emitted}$ is the number of emitted fluorescence photons and $N_{\rm abs}(\lambda_{\rm ex})$ is the number of absorbed excitation photons by the sample at the excitation wavelength $\lambda_{\rm ex}$.

The fluorescence of the sample is measured with a fluorescence spectrometer. The optical layout of the spectrometer is shown in Fig.~\ref{fig:FLQYSetup}. The light source is a Xenon lamp followed by a grating monochromator. The bis-MSB solution sample is contained in a 1 cm square quartz cuvette. Fluorescence is collected from the side of the cuvette, selected with another grating monochromator, and counted by a PMT.

$N_{\rm emitted}$ can be expressed~\cite{FLQY_Review} as
\begin{equation}
\label{relativeFLQY}
N_{\rm emitted}  =  \alpha \cdot  n^2\cdot I_0 \cdot \int s(\lambda)d\lambda \, ,
\end{equation}
where $s(\lambda)$ is the fluorescence intensity at wavelength $\lambda$ per unit excitation light intensity measured by the fluorescence spectrometer, $I_0$ is the excitation light intensity, $n$ is the refractive index of the sample, and $\alpha$ is a coefficient that converts the measured fluorescence intensity to emitted photon numbers. Due to the refraction of the light travelling from the liquid sample to the air, the acceptance depends on the refractive index of different samples\cite{n2corr_1952,n2corr_1970}. $\alpha$ includes geometric acceptance, PMT counting efficiencies, etc. In this study, we keep the absorption at the excitation wavelength smaller than 5\% at the cuvette center for each sample. The absorption along the light path in the cuvette is nearly uniform for all samples at wavelengths longer than the excitation wavelength. The geometric acceptance will be the same for these measurements and therefore $\alpha$ is almost a constant. At shorter wavelengths, absorption could be large for some measurements and the geometric acceptance may change. In this case, we will still keep the $\alpha$ fixed as determined by the calibration with a standard reference, and include the possible changes as an uncertainty.

\begin{center}
\includegraphics[width=8cm]{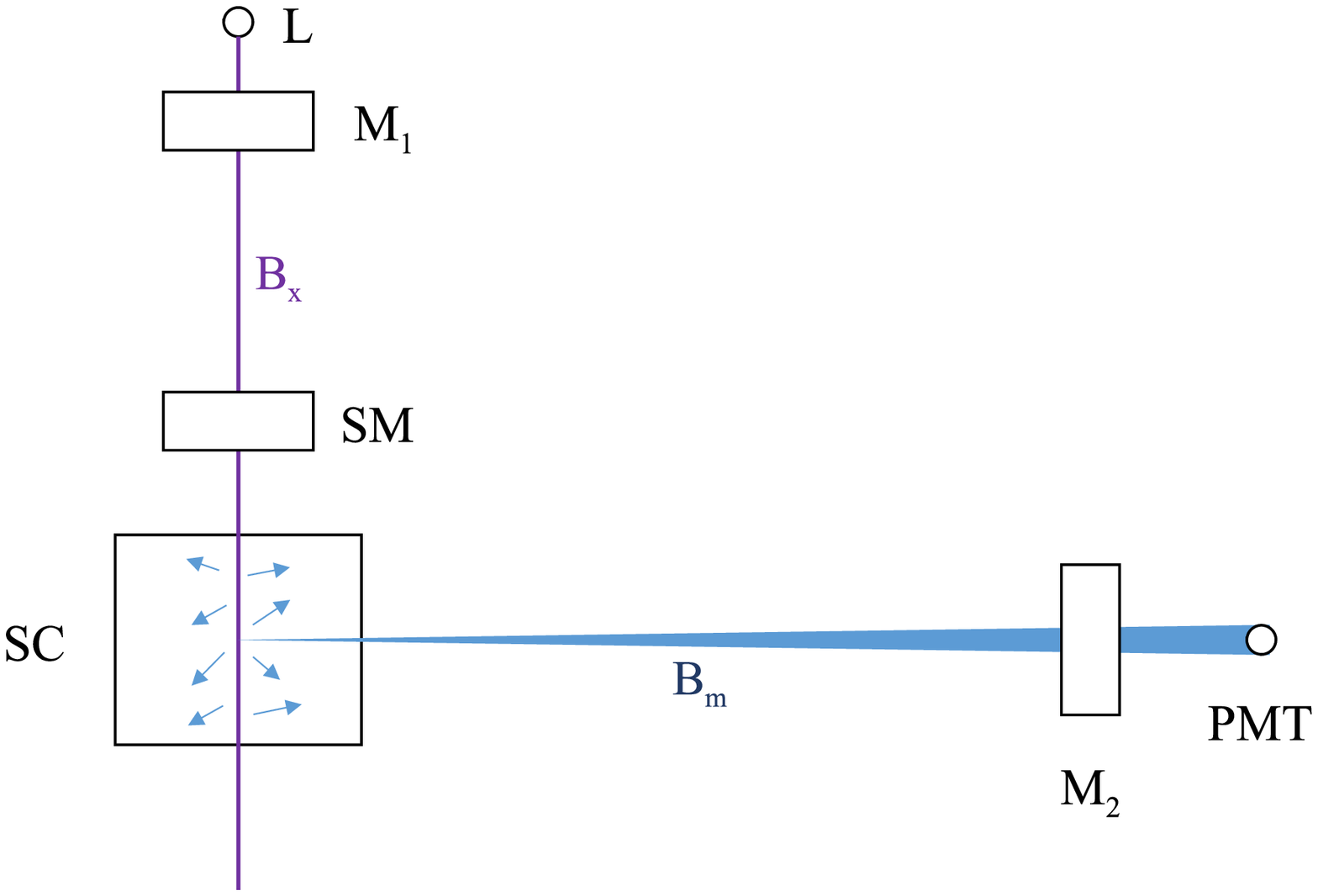}
\figcaption{\label{fig:FLQYSetup} Layout of the fluorescence spectrometer. B$_{\rm x}$, excitation beam; B$_{\rm m}$, emission beam; L, Xenon lamp; M$_1$, M$_2$ monochromators; SM, source monitor; SC, sample cuvette.}
\end{center}

The fluorescence emission is proportional to the light absorbed in the cuvette. The absorbance of the samples is measured with two UV-vis spectrometers.

We study the fluorescence quantum yield of bis-MSB by dissolving it in LAB. The absorption curve of the bis-MSB in LAB versus wavelength can be found in Fig.~\ref{fig:MolarExt}. At wavelengths longer than 400 nm, the absorption is very small. To extend the measurement to these wavelengths, we prepare multiple samples with different concentrations, from 0.082 mg/L to 150 mg/L. The fluorescence quantum yield should be independent of the concentration, unless the concentration is so high that the concentration quenching from collisions between solute molecules cannot be ignored. To compare the absorption at different concentrations, we express it as a molar extinction coefficient, assuming Lambert-Beer's law~\cite{BLlaw} is obeyed. The absorbance $A$ measured by the UV-vis spectrometer is
\begin{equation}
\label{BLlaw}
A = -\log_{10}(I/I_0) = \varepsilon\cdot c\cdot L \,,
\end{equation}
with $I_0$ the incident light intensity, $I$ the transmission light intensity, $\varepsilon$ the molar extinction coefficient, $c$ the concentration of bis-MSB, and $L$ the length of the cuvette.

\begin{center}
    \includegraphics[width=8cm]{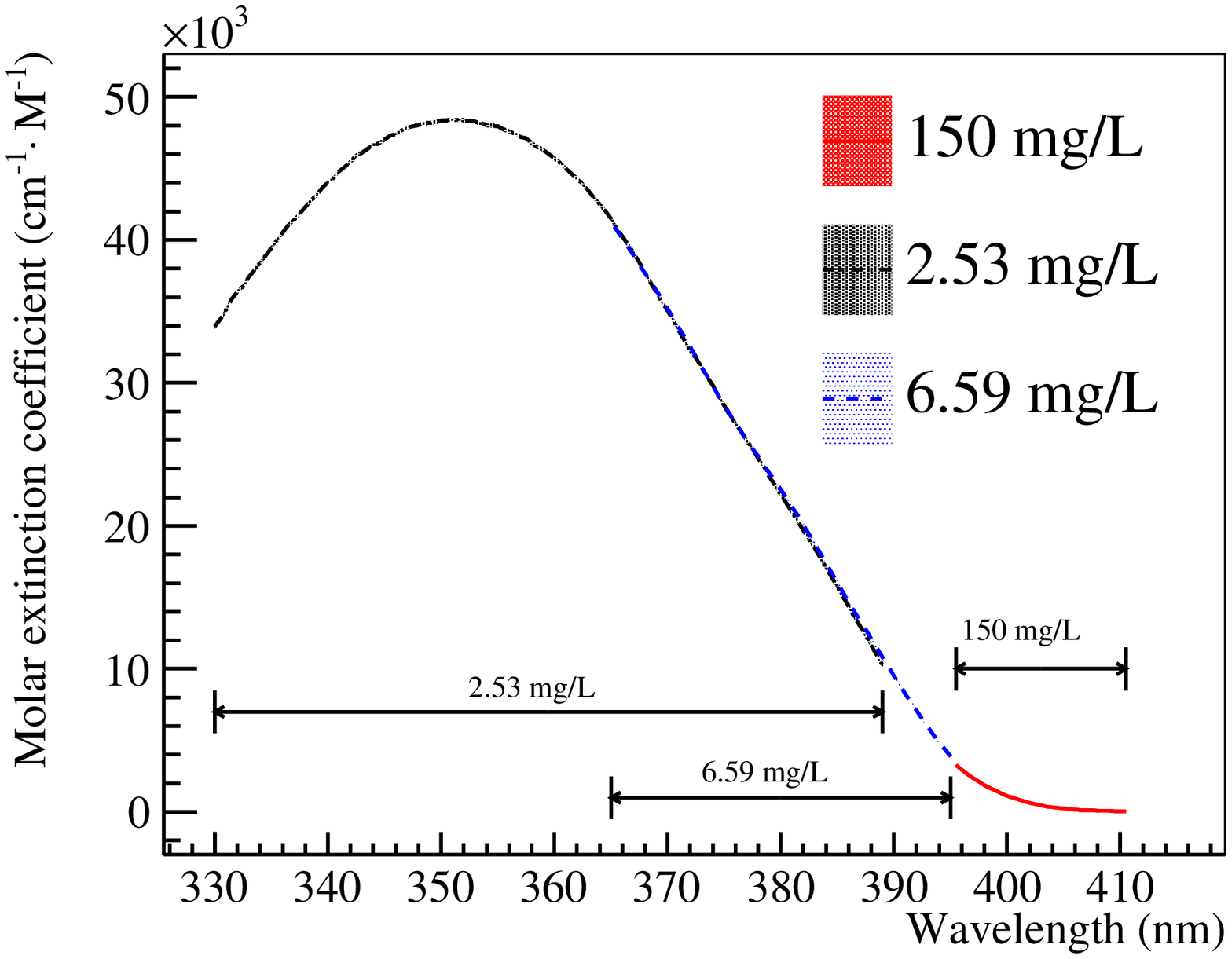}\\
\footnotesize{(a)}
  \hspace{1in}
    \includegraphics[width=8cm]{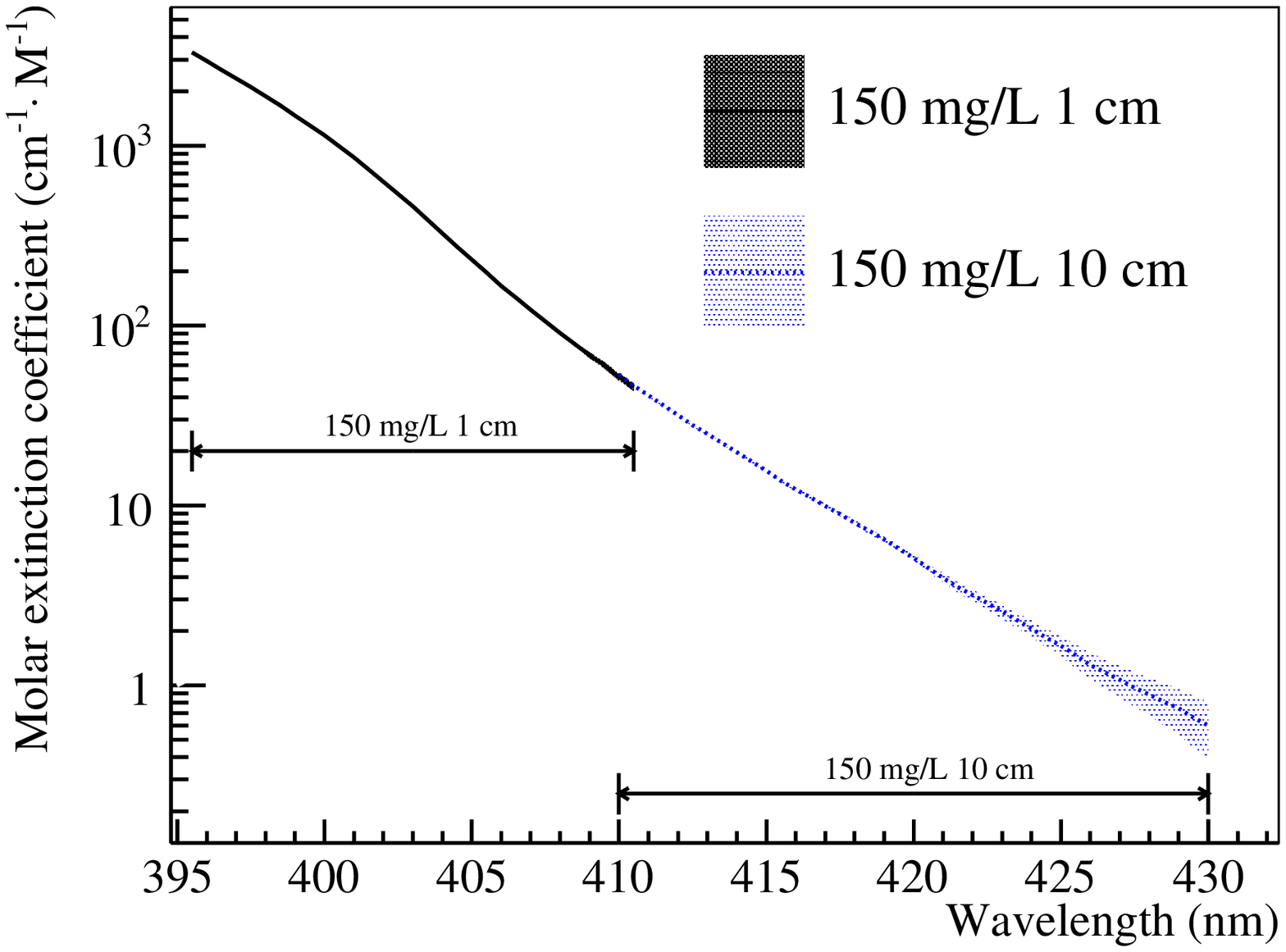}\\
\footnotesize{(b)}
  \figcaption{\label{fig:MolarExt} (a) Molar extinction coefficient of bis-MSB in LAB from 330 nm to 410 nm, using a 1cm cuvette for 3 different concentrations; (b) Molar extinction coefficient of bis-MSB from 395 nm to 430 nm, using 1cm and 10cm cuvettes for the 150 mg/L sample. The wavelength range for each measurement is shown on the plot.}
\end{center}

The absorption happens along the whole light path in the cuvette but the acceptance of the emitted fluorescence is not well known. In this study we assume that only the fluorescence emitted at the center of the cuvette can be observed by the spectrometer, and assign an uncertainty on the acceptance by assuming another extreme model where all the fluorescence emitted along the whole light path in the cuvette can be observed. The difference between the two models is actually small for our measurements. With this assumption, the number of absorbed photons in a narrow window around the cuvette center is $N_{\rm abs}\propto I_0^\prime-I$, where $I_0^\prime$ and $I$ are the light intensity entering and exiting this window, respectively. With the definition of the molar extinction coefficient in Eq.~(\ref{BLlaw}), it can be approximated as
\begin{equation}
\label{eq:nabs}
N_{\rm abs}\propto I_0^\prime \cdot \ln10 \cdot \varepsilon\cdot c \cdot d
\end{equation}
for a very small window length $d$. The light intensity $I_0^\prime=I_0\cdot 10^{-\varepsilon \cdot c\cdot L/2 - \epsilon\cdot L/2}$, where $I_0$ is the incident light intensity to the cuvette. The factor $10^{-\varepsilon \cdot c\cdot L/2}$ accounts for the light attenuation at the cuvette center due to the bis-MSB absorption, and $10^{-\epsilon\cdot L/2}$ accounts for that due to the LAB absorption. Since $L$ is not a small value, an approximation such as Eq.~(\ref{eq:nabs}) may introduce bias. The fluorescence quantum yield can be expressed as
\begin{eqnarray}
\label{ffeq}
\varphi(\lambda_{\rm ex}) = \alpha \cdot  n^2\cdot \dfrac{\int s(\lambda)d\lambda}
{\varepsilon \cdot c \cdot 10^{-\varepsilon \cdot c\cdot L/2 - \epsilon\cdot L/2}} \,,
\end{eqnarray}
where we have absorbed constants, such as the acceptance, $d$, and others, into the coefficient $\alpha$. The incident light intensity $I_0$ in $N_{\rm emitted}$ and in $N_{\rm abs}$ cancels out. Both $\varepsilon$ and $\epsilon$ are functions of the excitation wavelength $\lambda_{\rm ex}$.

Determination of the absolute efficiency of the measurement is difficult and may carry large uncertainty. The common practice is to calibrate the measurement with a standard reference sample~\cite{FLQY_Review}. In this study we use Harmane (CAS No. 486-84-0) solution, whose fluorescence quantum yield is well known as $\varphi_{\rm Harmane}=0.83\pm0.03$~\cite{Harmane}.

\section{Measurements and results}

The fluorescence quantum yield of bis-MSB was measured by dissolving it in specially purified LAB at different concentrations. The molar extinction coefficient $\varepsilon$ and the LAB absorption $\epsilon$ at different wavelengths was determined with the UV-vis spectrometers. The fluorescence yield $\int s(\lambda)d\lambda$ at given bis-MSB concentration $c$ was measured with the fluorescence spectrometer. The fluorescence quantum yield was then calibrated with the standard reference sample Harmane and calculated with  Eq.~\ref{ffeq}.

\subsection{Molar extinction coefficient}

A UV-vis spectrometer was extensively used in the Daya Bay experiment to measure and monitor the transparency of the liquid scintillator~\cite{DYBLS}. The uncertainty of the absorbance could reach 0.001 with the Daya Bay practices, although the specification of the used spectrometers, Shimadzu UV2550 and PerkinElmer L650, is $\sim\,$0.003, accounting for photometric accuracy, repeatability, flatness of the baseline, and stray light. We filled both the sample cuvette and the reference cuvette of the double-beam spectrometer with LAB to correct for the baseline, and then replaced the LAB in the sample cuvette with the bis-MSB solution for the absorbance measurement. In order to minimize possible systematic bias, all measurements were conducted with the same cuvettes under the same orientation. The uncertainty could be examined by comparing the measurements at two spectrometers and by checking the reproducibility. The absorbance at $\sim\,$600 nm, where both LAB and bis-MSB have little absorption, should be around zero, which provides another check on the uncertainty.

The LAB absorbance was measured versus an empty cuvette, after correcting for the baseline with two empty cuvettes. The absorbance was then normalized according to the attenuation length measurements with a 1-meter long tube~\cite{gaolong}, which showed that the attenuation length of the purified LAB is $\sim\,$20 m at 430 nm, corresponding to an absorbance $\sim\,$0.002.

The measured molar extinction coefficient of the bis-MSB is shown in Fig.~\ref{fig:MolarExt}. The uncertainty was propagated according to Eq.~(\ref{BLlaw}) from the uncertainty of the absorbance, which was taken as 0.001 as described above. The optimal absorbance range for an accurate measurement with the UV-vis spectrometer is between 0.2 and 0.8. Beyond this range, the uncertainty will increase. Therefore, multiple samples with different concentrations were prepared to account for the dramatic changes in absorbance and cover the whole wavelength range of interest. All the samples were measured in the same 1cm cuvette except for the 410-430 nm range with 150 mg/L sample, which has small absorption and thus was measured in a 10cm cuvette. The consistency of the different samples shows the reliability of the measurements and the validity of Lambert-Beer's law.

\subsection{Fluorescence intensity}

The fluorescence yield was measured with a fluorescence spectrometer HITACHI F4500. The layout of the spectrometer is shown in Fig.~1.

The fluorescence was counted by a PMT following a monochromator inside the spectrometer. The monochromator, as well as the one inside the light source, has a wavelength accuracy of $\pm 2$ nm and a resolution of 1 nm from its specification. The PMT dark count was recorded by the spectrometer for each measurement and subtracted. The measured emission spectrum was corrected for the PMT response at different wavelengths by the software provided by the manufacturer. Due to the residual PMT dark counts and stray light, the baseline was not flat. We measured the profile of the baseline with the empty cuvette as a template of the baseline for the emission spectrum fitting, with a free normalization parameter to allow scaling.

Two fluorescence spectra are shown in Fig.~\ref{fig:longHshort} with excitation wavelengths at 355 nm and 410 nm. The bis-MSB emissions spectrum ranges from 360 nm to $\sim\,$600 nm. Rayleigh scattering peaks are prominent at the excitation wavelengths, with a full width at half maximum (FWHM) of 5 nm. This is consistent with the specification of the spectrometer that the monochromator bandpass for both the excitation light and the emission light is 2.5 nm. The Rayleigh scattering light is mixed with the fluorescence when the excitation wavelength is 410 nm. The 410 nm excitation light can produce shorter wavelength fluorescence due to the interaction between the solute and solvent molecules, which causes rotation of the molecules, as well as the vibration of the molecules. The same mechanisms explain  why we actually observe a wide spectrum for the fluorescence instead of a single line when the molecules de-excite from the first excitation state to the ground state. Both states are broadened by these effects.

\begin{center}
\includegraphics[width=8cm]{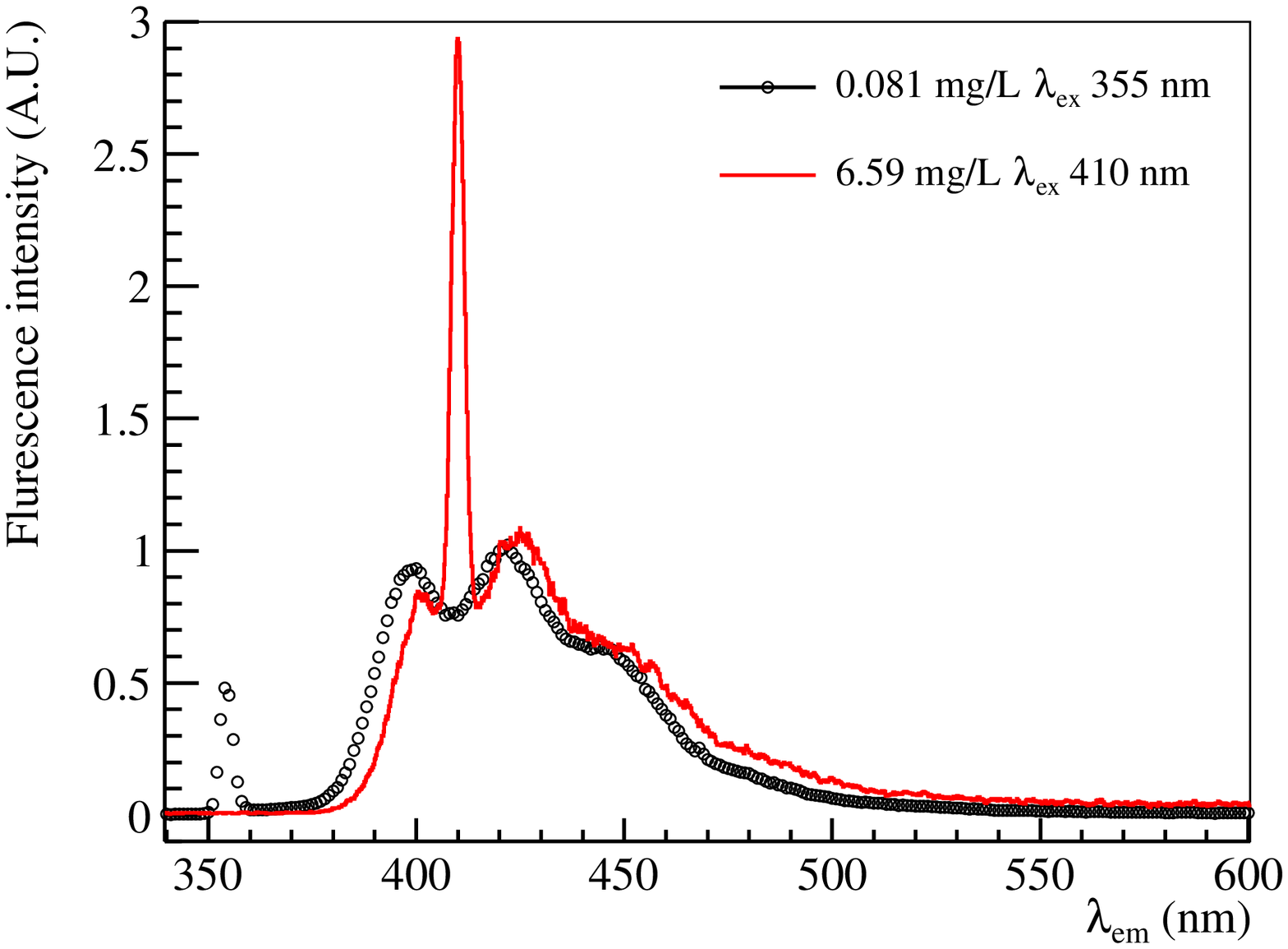}
\figcaption{\label{fig:longHshort} Bis-MSB fluorescence spectra with excitation wavelengths at 355 nm (black open circles) and 410 nm (red solid line). Rayleigh scattering peaks are prominent at the excitation wavelengthes. Samples of different concentrations, 0.081 mg/L and 6.59 mg/L, are used to keep the absorption at the excitation wavelength in a reasonable range. The fluorescence intensities are scaled arbitrarily.}
\end{center}

A shift and distortion in the spectrum is seen for these two measurements. This redshift of the emission spectrum mainly comes from the so-called ``red-edge" effect~\cite{RedEdgeReview}, happening when the excitation wavelength is on the red edge of the absorption band. Besides, the fluorescence at short wavelengths is absorbed and re-emitted at longer wavelengths, which also shifts and distorts the emission spectrum.

Rayleigh scattering needs be removed to obtain the fluorescence intensity. Since the Rayleigh scattering peak is large comparing to the fluorescence, subtracting it with a fit to the peak may introduce large uncertainty to the emission spectrum. Instead, we fitted the emission spectrum adjacent to the Rayleigh scattering peak to a template emission spectrum measured with the excitation wavelength well separated from the emission spectrum. The fitting was done with two parameters describing the shift along the x direction (wavelength) and the scale along the y direction (intensity), in a range of [-15,-5] nm and [5,15] nm around the excitation wavelength. The fluorescence emission spectrum was then obtained by replacing the measured spectrum with the best fit spectrum in a range of $\pm$5 nm around the excitation wavelength.

An example of the fluorescence spectrum fitting is shown in Fig.~\ref{fig:FitRayleigh}, in logarithmic scale with an excitation wavelength of 400 nm, where the black dots are the measured raw spectrum, the blue dotted line is the fitted baseline, and the red dashed line is the fitted fluorescence emission spectrum with Rayleigh scattering subtracted.

\begin{center}
\includegraphics[width=8cm]{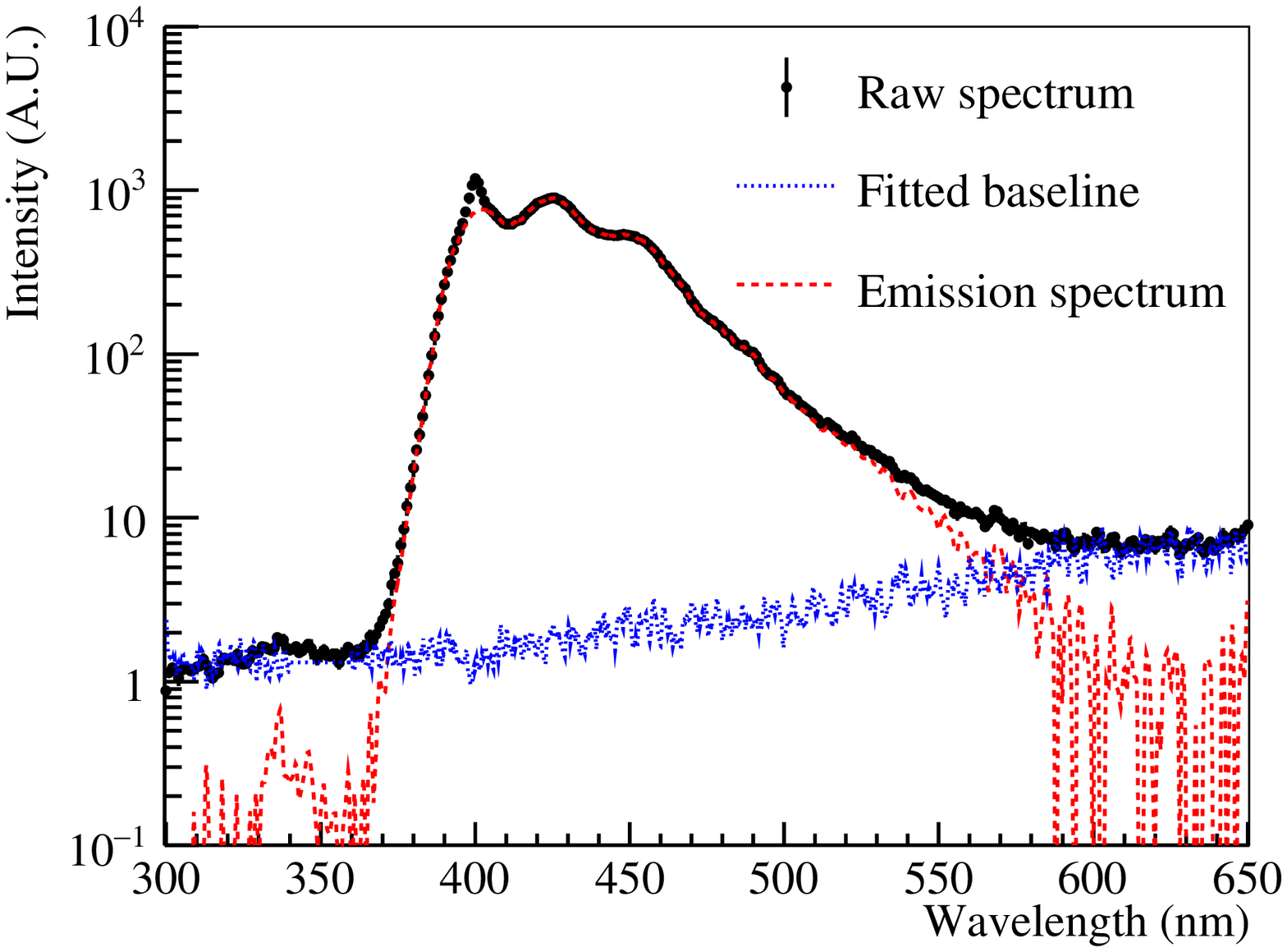}
\figcaption{\label{fig:FitRayleigh} A fluorescence spectrum measured with an excitation wavelength of 400 nm. The raw spectrum is shown as the black dots. The fitted baseline is shown as the blue dotted line. The fitted emission spectrum with the Rayleigh scattering peak subtracted is shown as the red dashed line.}
\end{center}

The emitted fluorescence can be absorbed in the liquid in the process of self-absorption, which reduces the light observed by the spectrometer. The absorption can  be roughly estimated with the measured molar extinction coefficient. However, since the acceptance of the emitted fluorescence is not well known, assuming the observed fluorescence comes from the center of the cuvette only, rather than from the whole light path along the cuvette, results in different self-absorption. As described in Sec.~2, we kept the absorption at the excitation wavelength smaller than 5\% at the cuvette center. The light path is 0.5 cm. The self-absorption for light on the way to the PMT is therefore also smaller than 5\% when the emission wavelength is longer than the excitation wavelength. The self-absorption distorts the emission spectrum if the absorption becomes large at short emission wavelength. Another estimation can be done by comparing the measured emission spectrum with that at relatively low concentration. We may observe the distortion at short wavelength. Such estimation is again not accurate since the emission spectrum could change at different excitation wavelengths. In this work, the self-absorption effect was not corrected for, but considered in the systematic uncertainty estimation to cover different models.

The emission light intensity, defined as the integrated emission light over the emission wavelength per unit concentration $\int s(\lambda)d\lambda/c$, is shown in Fig.~\ref{fig:IntUc}.
To ensure the absorption is smaller than 5\%, the measurements have been done at different bis-MSB concentrations. For multiple measurements at the same excitation wavelength, the excitation light intensity at the cuvette center differs, by at most 5\%. To be consistent for different concentrations, the intensity should be corrected by a factor $10^{-\varepsilon \cdot c\cdot L/2 - \epsilon\cdot L/2}$, but this was not done here. The light intensity drops 5 orders from 345 nm to 430 nm, reflecting the dramatic decrease of the molar extinction coefficient at long wavelength. Four uncertainties have been considered and included for the data points.

\begin{itemize}
  \item Uncertainty from the baseline subtraction. The baseline was measured with the empty cuvette but this does not fit the sample measurement. Scaling was allowed to fit the spectrum. We include 100\% uncertainty for the area of the baseline in the signal region (from 360 nm to 600 nm). This uncertainty is smaller than 3\% except that at the longest excitation wavelength for each sample, which was about 10\%.
  \item Uncertainty from the removal of Rayleigh scattering. The fitting error of the scaling factor of the emission template was taken as the uncertainty, and is smaller than 0.5\% for all data points.
  \item Uncertainty from the self-absorption effect. The self-absorption uncertainty was estimated as the difference between the undistorted emission spectrum (template) and the measured emission spectrum. For dilute samples (0.72 mg/L and 6.59 mg/L samples in this paper), the emission spectrum can be restored with the self-absorption correction with the measured absorbance. For concentrated samples (150 mg/L and 500 mg/L samples), the absorption is too strong and the emission spectrum cannot be restored with such a method. The excitation wavelengths we used for these samples are longer than some of the fluorescence. The emission spectra are distorted. However, it is reasonable to assume that the fraction of the emission spectrum integral of the short wavelengths (from 360 nm to 460 nm) in the whole emission spectrum (from 360 nm to 600 nm) are smaller than that of the dilute samples. This upper limit is taken as the uncertainty. The self-absorption uncertainty is smaller than 5\% for the dilute samples, and is large for concentrated samples. For example, the uncertainty of the 150 mg/L sample at 410 nm and the 500 mg/L sample at 415 nm are 19\% and 31\%, respectively.
  \item Uncertainty from the reproducibility of the fluorescence spectrometer, which is smaller than 0.5\%.
\end{itemize}

\begin{center}
\includegraphics[width=8cm]{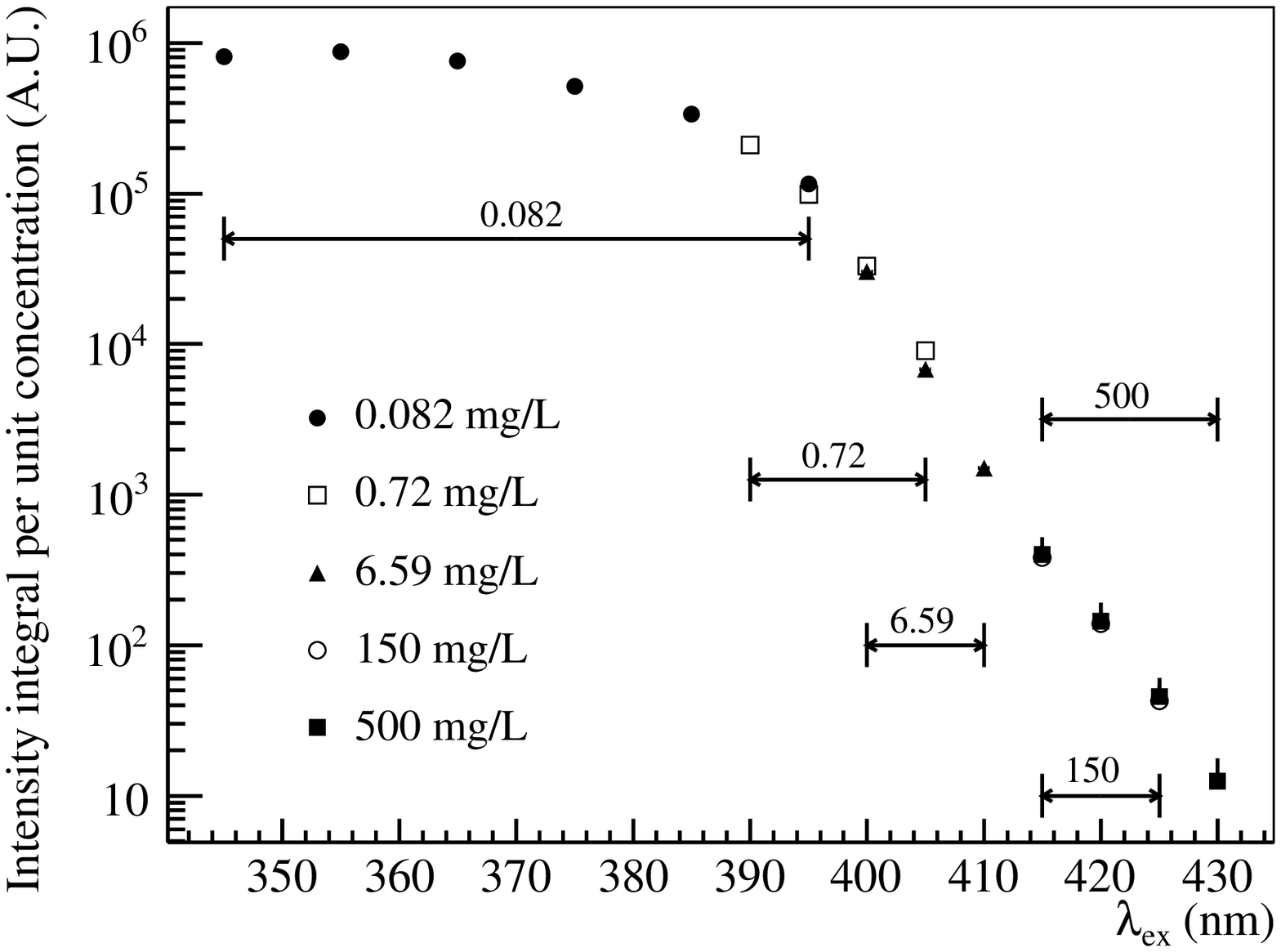}
\figcaption{\label{fig:IntUc} The integral of emission intensity per unit concentration.}
\end{center}

\subsection{Absolute efficiency}
The fluorescence standard Harmane was used to determine the absolute efficiency $\alpha$ in Eq.~(\ref{relativeFLQY}), with $\varphi=0.83\pm0.03$ for Harmane~\cite{Harmane}. The Harmane is dissolved in 0.1 N dilute sulphuric acid at 1.94 $\mu$~mol/L. All the measurement conditions are the same as the sample measurements. When calibrating the measurements for bis-MSB in LAB, the difference of the refractive index, 1.33 for the Harmane solution and 1.51 for the bis-MSB in LAB, has been corrected for as shown in Eq.~\ref{ffeq}. The variation of the refractive index at different wavelengths is negligible.

\subsection{Fluorescence quantum yield of bis-MSB}

The fluorescence quantum yield curve can be derived according to Eq.~(\ref{ffeq}), with the emission intensity per unit concentration shown in Fig.~\ref{fig:IntUc}, the molar extinction coefficient shown in Fig.~\ref{fig:MolarExt}, and the absolute efficiency $\alpha$ determined with the standard reference Harmane. The attenuation factor of the extinction light at the cuvette center has been included. Multiple measurements at different concentrations have been weight-averaged with their corresponding uncertainties. The obtained fluorescence quantum yield of bis-MSB is shown in Fig.~\ref{fig:FLQY},  where the uncertainty of the scale factor $\alpha$ is not included. At $\lambda_{\rm ex}$ = 350 nm, considering the uncertainty of $\alpha$, it is 0.926$\pm$0.053, which is consistent with the results given in Ref.~\cite{Argon} (0.93) and Ref.~\cite{XiaoHL} (0.96$\pm$0.03).

Two major uncertainty components, the emission intensity and the acceptance, are shown at the bottom of Fig.~\ref{fig:FLQY}. When the excitation wavelength is longer than 400 nm, the uncertainty mainly comes from the emission intensity, in which the self-absorption dominates. This uncertainty could be improved with Monte Carlo simulation to model the self-absorption effects. When the excitation wavelength is longer than 420 nm, the absorption of bis-MSB is very small even for high concentrations, and the uncertainty of the acceptance and the molar extinction coefficient is not negligible.

For comparison, the fluorescence quantum yield measured by Xiao {\it et al.}~\cite{XiaoHL} and that of Harmane are shown as the shaded areas. The re-emission probability of the Daya Bay liquid scintillator is shown as the red open rhombuses, and is discussed in Section 4.

\begin{center}
\includegraphics[width=8cm]{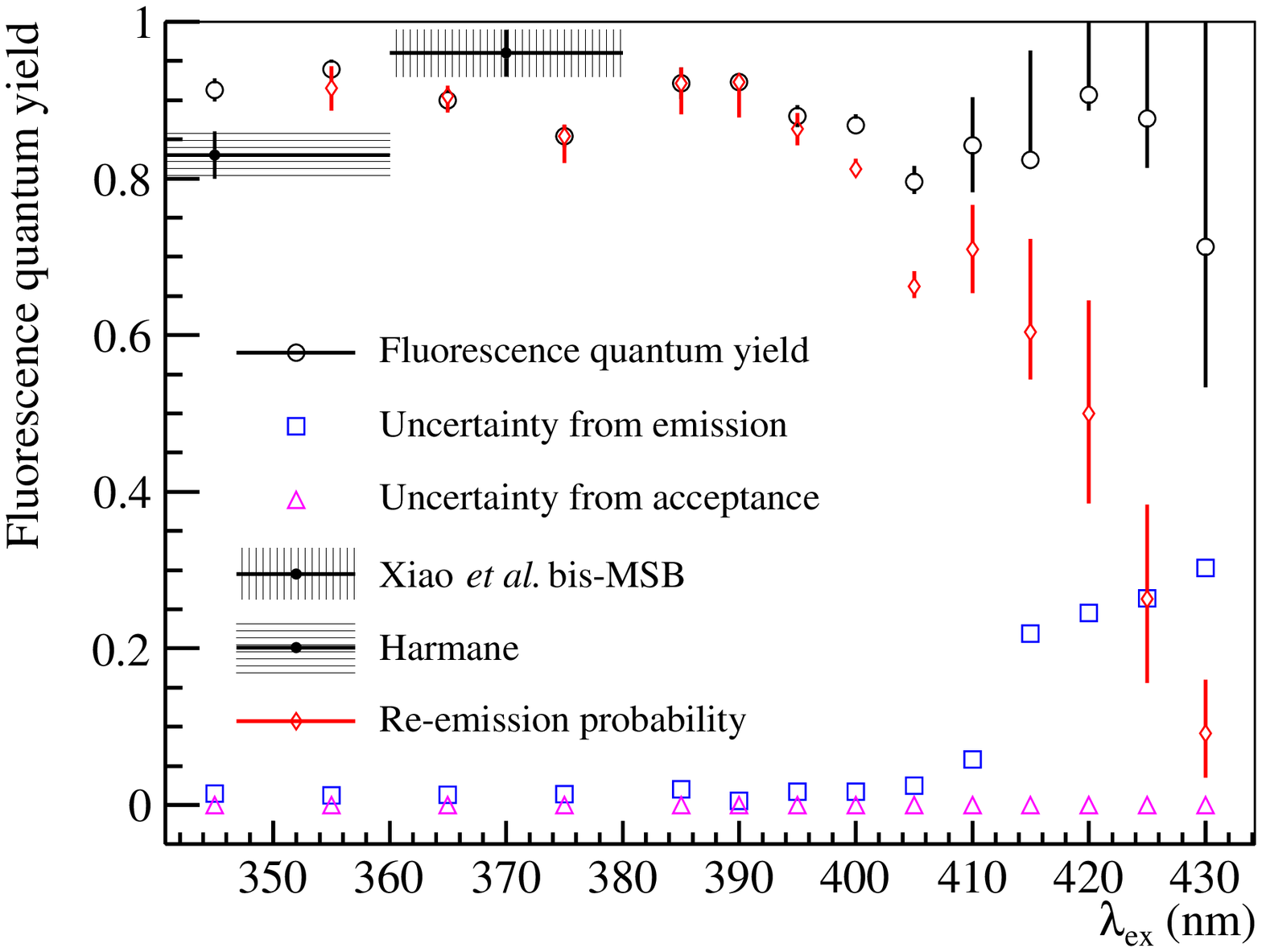}
\figcaption{\label{fig:FLQY} The fluorescence quantum yield of bis-MSB is shown as the black open circles. Two uncertainty components, the emission intensity and the acceptance, are shown as the blue open squares and the pink open triangles, respectively. The fluorescence quantum yield measured by Xiao {\it et al.} is shown as the shaded area filled with vertical lines, and that of the fluorescence standard Harmane is shown as the shaded area filled with horizontal lines. The re-emission probability of the Daya Bay liquid scintillator is shown as the red open rhombuses.}
\end{center}

\section{Discussion and conclusion}

We have measured the fluorescence quantum yield of bis-MSB from 345 nm to 430 nm. It is 0.926$\pm$0.053 at $\lambda_{\rm ex}$ = 350 nm where the absorption in the bis-MSB is strongest. The results are consistent with two previous measurements in literature but extend to much longer wavelengths, where the measurements are difficult due to the small absorption of bis-MSB, but important to understand the absorption re-emission process in the ternary liquid scintillator of the Daya Bay and JUNO experiments.

The re-emission probability is the probability that the photon is re-emitted after it is absorbed. Light absorption in the ternary liquid scintillator can happen in any of the three components. The re-emission probability not only depends on the fluorescence quantum yield of each component, but also depends on the absorption fraction of that component. At wavelengths longer than 400 nm, the fraction of the absorption by LAB molecules starts to increase but the absorbed light will not re-emit. The re-emission probability is calculated by modelling the absorption fraction of the liquid scintillator components and shown in Fig.~\ref{fig:FLQY}. The obtained fluorescence quantum yield of bis-MSB extends to the dropping edge of the absorption re-emission curve, which will enable us to model the optical processes precisely to predict the energy resolution of  huge liquid scintillator detectors like JUNO using Monte Carlo simulation.

\vspace{10mm}
\acknowledgments{We thank Ms. MA Li, Dr. WU Hai-Chen, and Dr. GU Zhan-Jun for their help and efforts. Ding X. F. thanks Ms. LI Le for the detailed discussion on the measurement.}

\end{multicols}

\begin{multicols}{2}

\end{multicols}

\clearpage

\end{document}